# Direct observation of heavy quasiparticles in the Kondo lattice CeIn$_3$

Y. Zhang,$^{1,2}$ W. Feng,$^1$ X. Lou,$^3$ T. L. Yu,$^3$ X. G. Zhu,$^1$ S. Y. Tan,$^1$ B. K. Yuan,$^1$ Y. Liu,$^1$ H. Y. Lu,$^1$ D. H. Xie,$^1$ Q. Liu,$^1$ W. Zhang,$^1$ X. B. Luo,$^1$ Y. B. Huang,$^4$ L. Z. Luo,$^1$ Z. J. Zhang,$^5$ X. C. Lai,$^{1\dagger}$ and Q. Y. Chen$^{1*}$

$^1$*Science and Technology on Surface Physics and Chemistry Laboratory, Mianyang 621907, China.*

$^2$*Department of Engineering Physics, Tsinghua University, Beijing 100084, China.*

$^3$*State Key Laboratory of Surface Physics and Department of Physics, Fudan University, Shanghai 200433, China.*

$^4$*Shanghai Institute of Applied Physics, Chinese Academy of Sciences, Shanghai, 201204, China.*

$^5$*Key Laboratory of Advanced Materials, School of Materials Science and Engineering, Tsinghua University, Beijing 100084, China.*

$^*$*Correspondence and requests for materials should be addressed to*

X. C. Lai $^\dagger$ (*email:* [laixinchun@caep.cn](mailto:laixinchun@caep.cn)) *and* Q. Y. Chen $^*$ (*email:* [chenqiuyun@caep.cn](mailto:chenqiuyun@caep.cn)).

The electronic structure of the Kondo lattice CeIn$_3$ has been studied by on-resonant angle-resolved photoemission spectroscopy and scanning tunneling microscopy/spectroscopy. A weakly dispersive quasiparticle band has been observed directly with an energy dispersion of 4 meV by photoemission, implying the existence of weak hybridization between the *f* electrons and conduction electrons. The hybridization is further confirmed by the formation of the hybridization gap revealed by temperature-dependent scanning tunneling spectroscopy. Moreover, we find the hybridization strength in CeIn$_3$ is much weaker than that in the more two-dimensional compounds CeCoIn$_5$ and CeIrIn$_5$. Our results may be essential for the complete microscopic understanding of this important compound and the related heavy-fermion systems.

## I. INTRODUCTION

An important issue concerning *d*-electron high temperature superconductors (HTSC) is how superconductivity (SC) emerges from magnetism, which is beyond the traditional Bardeen-Cooper-Schrieffer (BCS) framework and under intense debate.[1,2]

An alternative way to solve this problem is to understand the SC of *f*-electron heavy fermion (HF) systems, the behaviors of which are quite similar to the HTSC[3,4] and have close relationship with the *f*-electron properties.[5] In HF materials where localized *f* orbitals are arranged in a dense periodic array, the hybridization between those local moments with conduction electrons generates the composite quasiparticle with a heavy effective mass. Like other correlated electronic systems, such as HTSC, several of the HF compounds display an interplay between magnetism and SC and have a preference towards SC pairing with unconventional symmetry.[3,4] In these compounds, both magnetism and SC are originated from the *f* electrons, which are considered to have dual properties, i.e., both localized and itinerant character.[6] The localization of the *f* states promotes the formation of a magnetically ordered state, while the itinerancy of them favors the Fermi liquid state.[7] Moreover, when the localization and itinerancy of the *f* states are comparable, non-Fermi liquid behavior will appear and possibly the SC states.[8] Understanding the localized or itinerant character of the *f* electrons is quite important to unravel the ground-state properties and the relationship among *c-f* hybridization strength, magnetism and SC in HF and analogous *d*-electron systems.

In HF systems, the $Ce_nM_mIn_{3n+2m}$ (*M*=Co, Rh, Ir) family has rich phase diagrams and is a good target to address the complicated interactions in solids,[9-11] such as *c-f* hybridization, magnetism, SC, etc. Among them, $CeIn_3$ compound, being the parent material and fundamental unit of this family, has a cubic structure and orders antiferromagnetically below 10 K. The behaviors of the 4*f* electrons in $CeIn_3$ appears to be fairly important to understand the rich phase diagrams and different ground-state properties for the $Ce_nM_mIn_{3n+2m}$ family. However, previous measurements on the study of the *f*-electron properties of $CeIn_3$ did not reach a unanimous agreement. Transport,[12] magnetic,[13] optical conductivity,[14] and inelastic neutron scattering[15] results indicate the existence of HF state in $CeIn_3$ at low temperature. While soft x-ray photoemission,[16,17] angular correlation of the electron-positron annihilation radiation (ACAR)[18] and de Hass-van Alphen (dHvA)[19] measurements all demonstrate the localized character of the *f* states for $CeIn_3$. These contradicting results are partly rooted in the lack of understanding of the electronic structure of this important parent compound. Two very powerful tools which allow the study of these complex interactions and directly observe the behaviors of the $f$ electrons experimentally are angle-resolved photoemission spectroscopy (ARPES) and scanning tunneling microscopy/spectroscopy (STM/STS). In our previous study of CeIn$_3$ by soft x-ray ARPES,[16] we have provided the three-dimensional electronic structure of CeIn$_3$. However, due to the poor energy resolution (~100 meV) and small photoemission cross section for the 4$f$ states, subtle changes of the 4$f$ states are difficult to trace. Fortunately, on-resonant ARPES at the Ce 4$d$-4$f$ transition (121 eV) could largely enhance the $f$ electron photoemission matrix element, which has been proved to be an effective way to explore the $f$-electron states.[20-22] Furthermore, STM/STS is another effective approach to investigate the $f$-electron properties.[23] However, high resolution on-resonant ARPES and STM/STS studies are still lacking for this important parent HF compound CeIn$_3$.

Here the electronic structure of CeIn$_3$ is studied by combining on-resonant ARPES and STM/STS. We find a weakly dispersive quasiparticle band with an energy dispersion of 4 meV at the locations where $f$ band and conduction bands intersect by ARPES, indicating the hybridization between them. The hybridization is further supported by the formation of the hybridization gap revealed by temperature-dependent STS. Moreover, the hybridization strength between the $f$ electrons and conduction electrons in CeIn$_3$ is much weaker than that in the more two-dimensional CeCoIn$_5$ and CeIrIn$_5$ compounds.

## II. EXPERIMENT

High quality single crystals of CeIn$_3$ were grown by self-flux method. Soft x-ray ARPES data shown in Fig. 1(d) and Fig. 2(f) were obtained at the ADRESS end station of the Swiss Light Source facility. These spectra were taken using a PHOIBOS-150 photoelectron analyzer. The combined energy resolution is 80 meV or better and the angle resolution is 0.1°. The samples used in the soft-x ray ARPES with 590 eV photons and STM/STS measurements were obtained by cycles of Ar$^+$ ion sputtering and annealing.[16] All the data taken with 121 eV photons were obtained at the "Dreamline" beamline of the Shanghai Synchrotron Radiation Facility (SSRF) with a Scienta D80 analyzer. The energy resolution is 17 meV and the angle resolution is 0.2°. The samples were cleaved along the *c*-axis in ultrahigh vacuum at 13 K before performing ARPES measurements. The base pressures of the two systems are below $5\times10^{-11}$ mbar during the entire measurements. All ARPES measurements were performed at 13 K. STM measurements were performed with a low temperature ultrahigh vacuum system (base pressure, $1.2\times10^{-11}$ mbar). *dI/dV* curves were obtained simultaneously with the feedback loop off. *dI/dV* versus sample bias (*V*) was recorded by superimposing a small sinusoidal modulation (4 mV, 731 Hz) to the sample bias voltage, then the first-harmonic signal of the current was detected through a lock-in amplifier. An electrochemically etched tungsten tip was used, and the STM image in this work was recorded in a constant-current mode.

## III. RESULTS

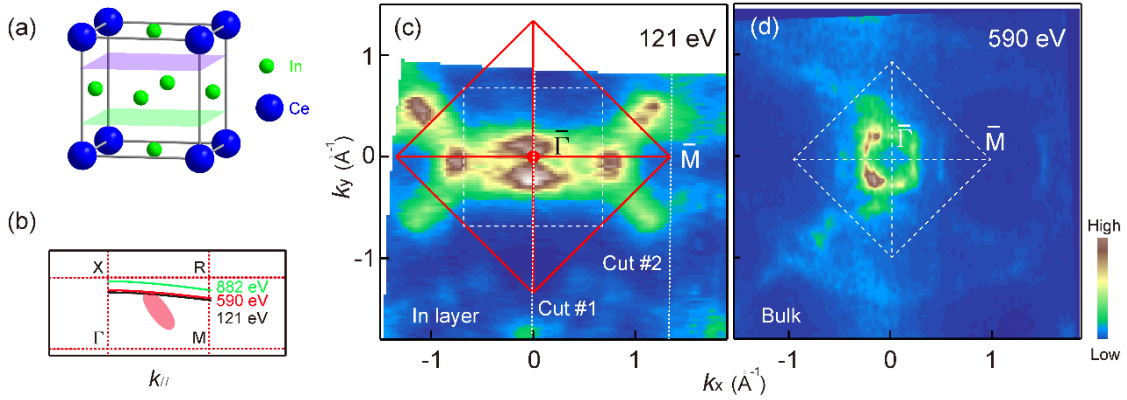

Figure 1. (a) Cubic crystal structure of $CeIn_3$ with a lattice constant of *a*=0.469 nm. The purple (green) plane indicates the In (Ce-In) layer-terminated surface, which exhibits an in-plane lattice constant of $a/\sqrt{2}$ =0.332 nm (*a*=0.469 nm). (b) The momentum cuts for photon energies of 121 eV, 590 eV and 882.5 eV are calculated with the inner potential of 15 eV, which is determined based on the periodicity of the high symmetry planes from soft x-ray ARPES data.[16,24] Here in order to indicate the relative $k_z$ positions with different photon energies to the heavy quasiparticles, we mark them in the same Brillouin zone (BZ). The purple shadow in the inset represents the calculated location of heavy quasiparticles.[24,25] (c) On-resonant photoemission intensity map at $E_F$ integrated over a window [$E_F$-15 meV, $E_F$+15 meV] with linear vertical (LV) polarized light for In layer-terminated $CeIn_3$ (001) surface. (d) Soft x-ray

photoemission intensity map at $E_F$ integrated over a window [$E_F$-100 meV, $E_F$+100 meV] with LV polarized light for Ce-In layer-terminated surface. The solid red in (c) and white dashed squares in (c) and (d) represent the projected BZ calculated with the in-plane lattice constants of In and Ce-In layer-terminated surfaces, respectively.

Figures 1(a-c) show the basic crystal and electronic structure of the cleaved surface of CeIn$_3$ by on-resonant ARPES with 121 eV photons. The momentum cut taken with 121 eV photons locates at approximately $k_z$=0.75 $\pi/a$ along Γ-X (M-R) as shown in Fig. 1(b). Here we define $k_z$=0 ($\pi/a$) for the Γ (X) point. The observed Fermi surface (FS) in Fig. 1(c) matches well with the projected BZ calculated with the lattice constant of the In layer-terminated surface of 0.332 nm. This value is much smaller than that of the Ce-In layer-terminated surface of 0.469 nm. Since CeIn$_3$ is cubic and hard to cleave, it is possible that small contributions from the Ce-In layer also exist. Here we can only confirm that the obtained spectra are dominated by In layer-terminated surface. On the other hand, the samples measured with 590 eV photons are obtained by cycles of sputtering and annealing method. The observed FS in Fig. 1(d) matches well with the projected BZ calculated with the lattice constant of Ce-In layer. Moreover, we have performed STM measurements (see Fig. 5) on the samples prepared by the same method, and only one termination is observed, the lattice constant of which is consistent with that of the Ce-In layer. For simplicity, we will refer to In (Ce-In) layer-terminated dominated surface as In (Ce-In) layer-terminated surface throughout the work.

The topology of the FS of the In layer-terminated surface in Fig. 1(c) consists of an elliptical-shaped electron pocket centred at the $\bar{\text{M}}$ point, a hole pocket and an electron pocket centred at the $\bar{\Gamma}$ point. For comparison, the electronic structure of the Ce-In layer-terminated surface at the same $k_z$ by soft x-ray ARPES is displayed in Fig. 1(d). The main features are similar to that in Fig. 1(c), except that the electron pocket centred at the $\bar{\Gamma}$ point is not that pronounced.

Previously, dHvA experiments, calculations and ACAR results[18,19,26,27] all indicate that there are three pronounced FS sheets for CeIn$_3$. One electron pocket is centred at the R point, the other two centred at the Γ point (one electron and one hole). Those results agree well with our ARPES data in Fig. 1(c) and 1(d). Since ARPES results with 590 eV photons are more bulk-sensitive than that with 121 eV photons, here we compare the size of the projected FS between the soft x-ray ARPES data in Fig. 1(d) and those from dHvA. Previous dHvA experiments and related calculation results[19,26,27] indicate that the dHvA frequency for the pressured paramagnetic (PM) CeIn$_3$ is about $F=10^4$ T ($S_F=0.595$ (1/Å)$^2$) for the electron pocket centred at the R point, while about $F=4\times10^3$ T ($S_F=0.238$ (1/Å)$^2$) for the hole pocket centred at the Γ point when magnetic field is along <100> direction. Here $F$ is the dHvA frequency and $S_F$ is the maximum or minimum cross-sectional area of the FS. CeIn$_3$ in the antiferromagnetic (AFM) state possesses FS sheets only 80-93% the size of the pressured one. From our ARPES results in Fig. 1(d), the size of the electron pocket around the $\bar{M}$ point is about 0.675 (1/Å)$^2$ and 0.24 (1/Å)$^2$ for the hole pocket around the $\bar{\Gamma}$ point, which are consistent with the dHvA results.

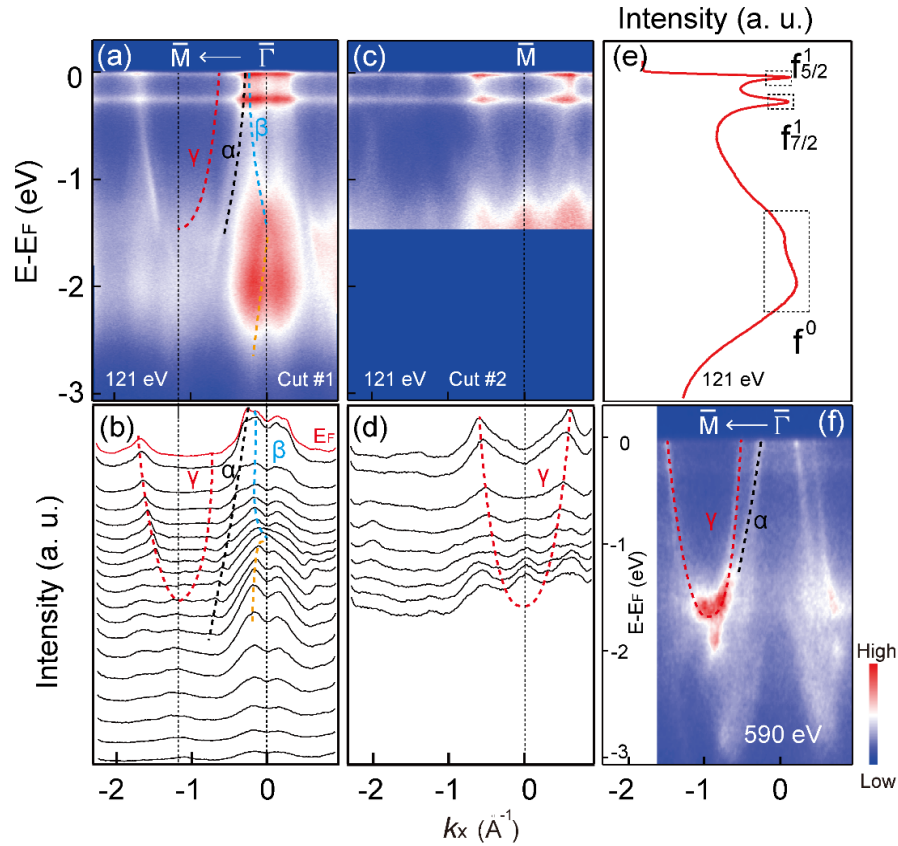

Figure 2. (a)-(b) On-resonant valence band structures of the In layer-terminated CeIn$_3$ (001) surface with LV polarized light along cut #1 labelled in Fig. 1(c) and the corresponding momentum distribution curves (MDCs). The color dashed lines in Fig. 2(a) are calculation results along the X-R direction,[28] where the Ce 4$f$ electrons are treated as localized core states. (c)-(d) On-resonant valence band structures of the In layer-terminated CeIn$_3$ (001) surface with LV polarized light along cut #2 labelled in Fig. 1(c) and the corresponding MDCs. The color dashed lines in Fig. 2(b) and (d) are guides to eyes. (e) Angle-integrated photoemission spectroscopy of the intensity plot in Fig. 2(a). (f) The electronic structure of the Ce-In layer-terminated surface along $\bar{\Gamma}$-$\bar{M}$ with LV polarized soft x-ray light. The color dashed lines are guides to eyes.

Figures 2(a)-(d) display the electronic structure of CeIn$_3$ taken with 121 eV photons. A pronounced feature of the on-resonant valence band structure is the weakly dispersive $f$ bands located at E$_F$ and 2 eV binding energy (BE), which correspond to the $f^1$ and $f^0$ states, respectively.[17] Two nearly flat bands located at E$_F$ and 260 meV BE can be clearly observed in Figs. 2(a) and (c), which can be assigned to the $4f^1_{5/2}$ state and its spin orbit coupling (SOC) sideband $4f^1_{7/2}$, respectively.[22,29] Additionally, two broad and flat bands located at 2 eV and 1.5 eV BE can also be observed in Fig. 2(a) and 2(e). The flat band at 2 eV BE arises from the pure charge excitations ($4f^1$-$4f^0$) and is usually referred to as the ionization peak.[17] The other broad band observed at 1.5 eV BE may reflect hybridization spreading due to structure in the valence band density of states, which has been observed in CeRh$_2$Si$_2$[30] and the monolayer of Ce film on W (110).[31] An electron-like band β, a hole-like band α centred at the $\bar{\Gamma}$ point and an electron-like band γ centred at the $\bar{M}$ point could also be clearly observed in Figs. 2(a)-(d). Those experimental band structures agree well with the calculation results[26,28] and are mainly derived from Ce 5$d$ and In 5$p$ orbitals. Since the β band is mainly derived from the In 5$p$ orbital and 121 eV photons are more surface-sensitive than 590 eV photons, more contributions from In 5$p$ orbitals may be observed from the 121 eV spectra on the In layer-terminated surface, which makes the β band much stronger in Figs. 1(a)-(b). This may be the reason that the β band is not that pronounced in Fig. 2(f).

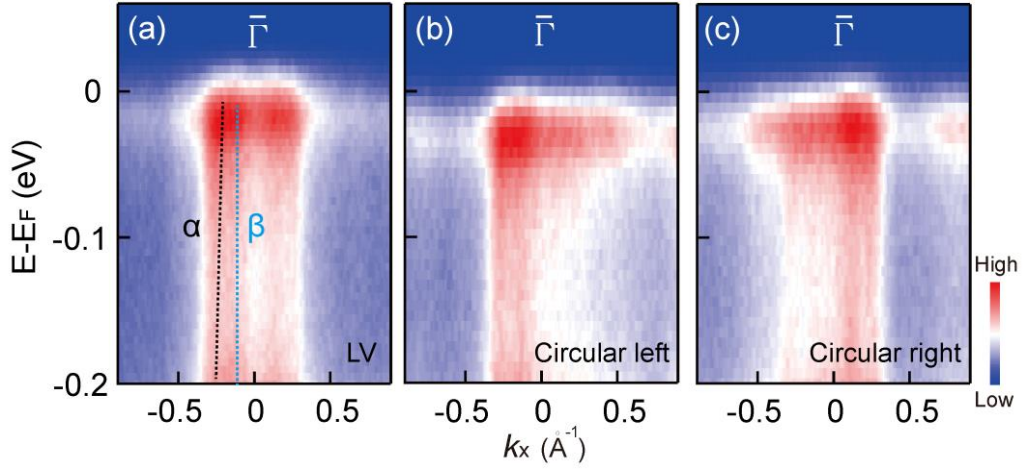

Figure 3. Photoemission intensity plots of the In layer-terminated surface of CeIn$_3$ along $\bar{\Gamma}$-$\bar{M}$ with 121 eV photons and (a) LV, (b) circular left, and (c) circular right polarization. The dashed lines in (a) are guides to eyes.

To resolve different orbital symmetry of the valence bands, on-resonant ARPES measurements with different polarized lights are performed, as shown in Fig. 3. In ARPES experiments, states with the same symmetry can be judged by their similar response to the change of the polarization. Moreover, matrix element effect for excitations of a photoelectron from the odd-symmetry states with circularly left or circularly right polarized light are affected by the possible admixture of the even-symmetry wave function in the states and vice versa. This will cause dichroic effects differing in sign for positive and negative $k$. States with the same dichroic response to a change of circular polarization can be assumed to have similar symmetry properties.[32] The dichroic effect in the band structure of CeIn$_3$ can be observed in Figs. 3(b)-(c), where ARPES data taken with left and right circular polarization are exhibited, respectively. Obviously, the β and α bands are both subject to dichroism near $E_F$. They are both largely enhanced on the left and right sides of the center $\bar{\Gamma}$ point upon the excitation with the left and right circularly polarization, suggesting the similar orbital symmetry of the β and α bands. Most importantly, the spectral intensity of the $4f_{5/2}^1$ state around the $\bar{\Gamma}$ point, shows dichroism similar to the β and α bands, indicating the same orbital symmetry between the three bands (β, α and $f$) and providing a platform for the formation of possible hybridization gap between them.

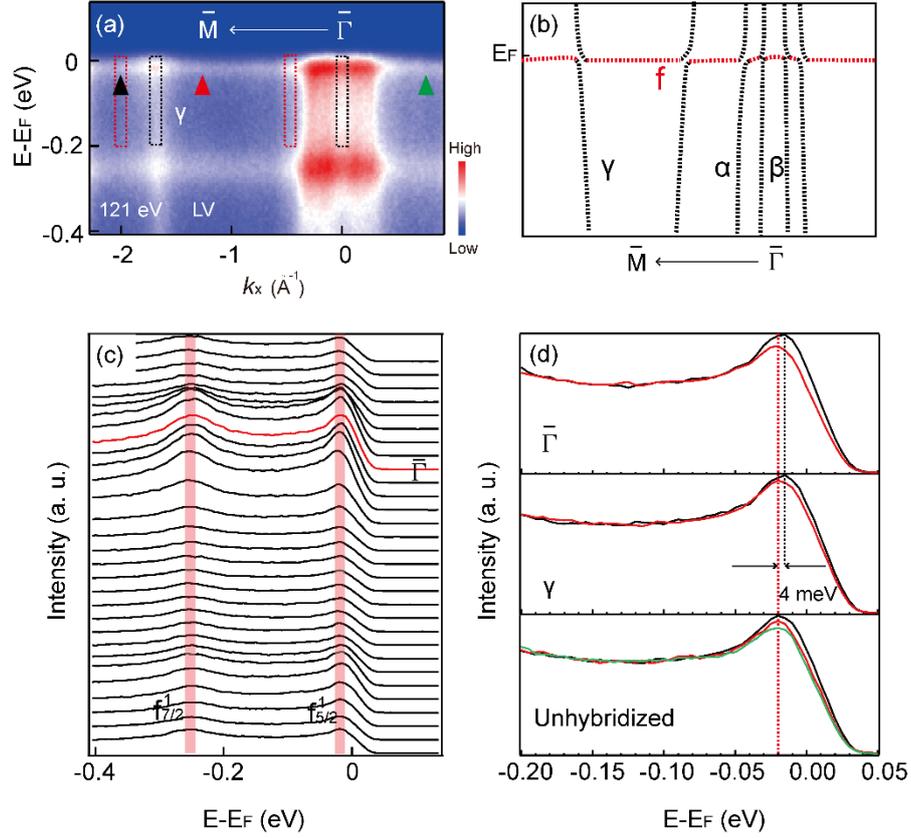

Figure 4. (a) On-resonant valence band structure of In layer-terminated CeIn$_3$ surface along $\bar{\Gamma}$-$\bar{M}$ with LV polarized light. (b) Schematic band structure of CeIn$_3$ extracted from Fig. 4(a). (c) Energy distribution curves (EDCs) of the photoemission intensity plot in Fig. 4(a). The pink shadows mark the positions of the $4f^1_{5/2}$ and $4f^1_{7/2}$ states, respectively. The red curve represents the EDC at the $\bar{\Gamma}$ point. (d) Comparisons of EDCs measured at different momentum locations in Fig. 4(a). The upper and middle panels correspond to the EDCs measured around the $\bar{\Gamma}$ point and the γ band, respectively. The black and red curves represent the integration ranges marked by black and red dashed rectangles in Fig. 4(a), respectively. The lower panel corresponds to the EDCs measured at the momentum locations marked by the colored triangles in Fig. 4(a).

For HF systems, due to the hybridization between the conduction band and the Ce 4$f$ state, a dispersive quasiparticle band could be observed around E$_F$ at the locations where $f$ band and conduction band intersect, which is schematically displayed in Fig. 4(b). The quasiparticle band has two features: strong intensity and a shift in BE at different momentum locations. Moreover, if the hybridization strength is strong enough, the intensity of the $4f^1_{7/2}$ state should be much weaker than that of the $4f^1_{5/2}$ state, as shown in CeCoGe$_{1.2}$Si$_{0.8}$.[20] Meanwhile, the BE shift of the quasiparticle band will become more obvious and much larger. We further focus on the electronic structure near E$_F$ to illustrate the $f$-state properties of CeIn$_3$ in Fig. 4. First, we notice that the intensity of the $4f^1_{7/2}$ state in CeIn$_3$ is comparable with that of the $4f^1_{5/2}$ state in Fig. 4(a) and 4(c), implying the highly localized nature of the $f$ electrons. This phenomenon has been observed by the soft x-ray ARPES in our previous work.[16] Interestingly, however, we find that the spectral intensities of both $4f^1_{5/2}$ and $4f^1_{7/2}$ states exhibit obvious momentum dependence, which can be revealed in Figs. 4(a) and 4(c) and were not observed in our previous work. The spectra of the $f$ states centred at the $\bar{\Gamma}$ point show strong intensity, so does the $f$ spectra at the intersection point of the γ and $f$ bands centred at the $\bar{M}$ point. While at other momentum locations, where there are no Fermi crossings of the valence bands, the spectral intensity of the $f$ states seems homogeneous and relatively weak.

Figure 4(d) enlarges the energy distribution curves (EDCs) of the $f$ spectra at different momentum locations. Around the $\bar{\Gamma}$ point, where the highest intensity locates, we find that the $f$ band exhibits a slight dispersion in the upper panel in Fig. 4(d). The existence of an energy dispersion and the enhanced intensity of the $f$ band around the $\bar{\Gamma}$ point indicate the possible formation of a quasiparticle band due to the ongoing hybridization between the $f$ electrons and conduction electrons. Similar weakly dispersive quasiparticle band can be observed at the intersection point of the $f$ band and γ band in the middle panel in Fig. 4(d). Especially, the hybridization emerges at the locations where the conduction bands approach or cross the Fermi level. While at other momentum locations, where no conduction bands cross the Fermi level, the $f$ band doesn't show observable energy dispersions in the lower panel in Fig. 4(d) accompanied with weak spectral intensities. Those results agree well with the calculations[26], where bands α, β, γ exhibit obvious hybridizations with $f$ bands.

By comparing the quasiparticle band peak positions at different momentum locations, we find that the energy dispersion of the hybridized band is about 4 meV. For

HF systems, the band structures near the Fermi level can be described by a mean-field hybridization band picture based on the periodic Anderson model.[33] According to the periodic Anderson model, larger hybridization strength between the conduction electrons and $f$ electrons induces a larger BE shift in the $f$-level band structure. If the hybridization strength is zero, the $f$ band doesn't show any energy dispersion in the band structures. For other prototypical HF compounds like $CeCoIn_5$ and $CeIrIn_5$,[22,29] the energy dispersions of the quasiparticle bands are more than 10 meV. This means that the hybridization strength between $f$ electrons and conduction electrons for $CeIn_3$ is rather weak, compared with $CeCoIn_5$ and $CeIrIn_5$.

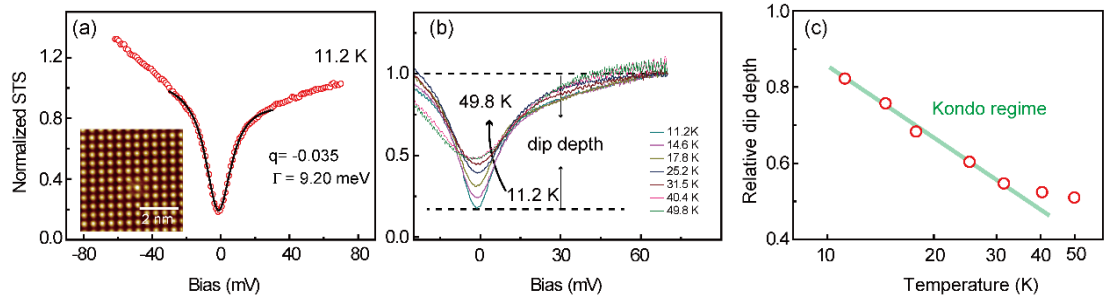

Figure 5. (a) $dI/dV$ spectrum (red dots) on Ce-In layer-terminated surface of $CeIn_3$ at 11.2 K. The black curve is the Fano fitting of the raw data (red dots). The inset is the corresponding atomic resolution image of $CeIn_3$. (b) Temperature evolution of the $dI/dV$ spectra up to 49.8 K. All spectra are normalized to the values at $V$= 70 mV. Here we define "dip depth" as the difference between the $dI/dV$ values at $V$=70 mV and $V$=0 mV. The dip depth at 11.2 K is marked by the two dashed lines. (c) The evolution of the Kondo dip depth as a function of temperature. The green line is a signature of a logarithmic fit.

To further confirm the formation of heavy quasiparticles in $CeIn_3$ at low temperature, STM/STS experiments were performed to detect the electronic structure of $CeIn_3$. The atomic resolution image in the inset of Fig. 5(a) shows that the atomic spacing is 0.468 nm, which is consistent with the lattice constant of the Ce-In layer-terminated surface of $CeIn_3$. Furthermore, a V-shaped gap feature around $E_F$ is observed from the $dI/dV$ curve of $CeIn_3$ in Fig. 5(a), which can be well fitted with a Fano spectrum.[34] Our STS results on Ce-In terminated surface of $CeIn_3$ are similar to that of the Ce-In terminated surface of $CeCoIn_5$.[23] In a Kondo system, the Fano line shape naturally occurs because of the presence of two interfering tunneling paths from STM tip, one directly into the itinerant electrons, and the other indirectly through the heavy quasiparticles.[35] The Fano resonance line shape follows:

$$dI/dV \propto \frac{(\varepsilon+q)^2}{1+\varepsilon^2}, \quad \varepsilon = \frac{eV-\varepsilon_0}{\Gamma},$$

Here $q$ reflects the quality of the ratio of probabilities between the two tunneling paths, $\varepsilon_0$ is the energy location of the resonance, and $\Gamma$ is the resonance half width at the half maximum (HWHM). This gap feature is related to the hybridization between $f$ electrons and conduction electrons and widely observed in other HF systems.[23,35,36] By fitting, we can obtain the parameters of the Fano curve at 11.2 K: $q$=-0.035 and $\Gamma$=9.2 meV, as shown in Fig. 5(a). The gap value in $CeIn_3$ from above fitting is about 18.4 meV, in agreement with the optical conductivity spectra result of $CeIn_3$, which reveals a hybridization gap of ~ 20 meV.[14]

In contrast, based on our ARPES measurements in Fig. 4, a hybridization gap of 4 meV can be obtained. The gap value obtained by our ARPES measurements is much smaller than that of the STS results. There are mainly two reasons responsible for this: 1) For Kondo lattice systems, the hybridization between the $f$ electrons and conduction electrons results in two separate bands, which gives a direct hybridization gap and a much smaller indirect gap.[23] The 4 meV gap revealed by ARPES in our experiments is the indirect gap, while the observed gap of 18.4 meV by STS is the direct gap. 2) The hybridization strength between the $f$ bands and different conduction bands in HF compounds may be different, which results in different values of the hybridization gap. ARPES measurements can provide momentum information, while STS can only give total density of states. Consequently, it is not that straight to compare the hybridization gaps between STM and ARPES results directly. Different values of hybridization gap from ARPES and STS have also been observed in $CeCoIn_5$.[22,23]

When the temperature is increased, we find that the gap feature becomes shallower in Fig. 5(b). Following previous analysis,[36] we obtain the relative dip depth as a function of temperature in Fig. 5(c). The dip depth can be reasonably well described by a logarithmic temperature dependence below 40 K, which is similar to other Kondo systems,[36-38] further indicating the emergence of HF state in $CeIn_3$. Such behavior is consistent with the temperature dependence of the resistivity of CeIn$_3$, which shows a maximum at 50 K,[13] indicating the formation of a coherent state. In addition, we find that the Kondo dip depth of CeIn$_3$ starts to increases below 40 K, while the depth of CeCoIn$_5$ already increases below 60 K.[23] This also implies weaker hybridization strength for CeIn$_3$ than CeCoIn$_5$.

## V. DISCUSSION AND CONCLUSIONS

In the Doniach phase diagram, the intersite Ruderman-Kittel-Kasuya-Yoshida (RKKY) interaction and onsite Kondo effect are competing and the Kondo effect could exist inside the AFM state. For CeIn$_3$ compound, the temperature-dependent electrical transport, heat transport and magnetic susceptibility curves[12,13] all exhibit a coherent-incoherent crossover behavior below 50 K, indicating the emergence of HF state at low temperature. These results are confirmed by the optical conductivity,[14] inelastic neutron scattering[15] results and the largely enhanced Sommerfield coefficient extracted from the specific heat measurements of CeIn$_3$.[11] However, our previous soft x-ray ARPES study of CeIn$_3$ fails to trace the signal of hybridization between the $f$ electrons and conduction electrons.[16] We propose that there are two main reasons responsible for the absence of the quasiparticle band in the soft x-ray band structures. i) As the dispersion of the quasiparticle band of CeIn$_3$ is no more than 4 meV, it is difficult for the soft x-ray ARPES measurements with a poor energy resolution of 80 meV to detect this subtle change. ii) According to the calculations,[24,25] the electronic structure of CeIn$_3$ is quite three-dimensional. The heavy quasiparticles are mainly centred at the momentum locations $K=<k,k,k>$, where $k=(0.5 \pm 0.1)\pi$. The momentum cut with 121 eV photons is closer to the $k_z$ positions where the quasiparticle bands locate in the BZ than that with 882.5 eV photons, as shown in Fig. 1(b).

We directly observe the dual properties of the $f$ states in CeIn$_3$ at low temperature. Most of the $f$ electrons stay localized, while a small portion of $f$ electrons participate in the formation of FS. This situation will change if we add a positive pressure on CeIn$_3$ crystals. dHvA experiments and theoretical results[19,39] manifest that a positive pressure on CeIn$_3$ indicates enhanced hybridization strength between $f$ electrons and conduction electrons. More $f$ electrons start to participate in the modification of FS and the collective behaviors of the $f$ electrons make this system itinerant and heavy. At the same time, the ground state of $CeIn_3$ will also change to the SC state or PM Fermi liquid state under pressure.[12] Those phenomena give a clear insight into the relationship between the hybridization strength and ground-state properties. We further introduce $CeMIn_5$ (*M*=Co, Rh, Ir) compounds for comparison. As layered compounds, the structure of $CeMIn_5$ is comprised of alternating layers of $CeIn_3$ and $MIn_2$ and has a two-dimensional feature. The three-dimensional component $CeIn_3$ layer in $CeMIn_5$ compounds contributes all the $f$ electrons and can be viewed as adding an effective positive pressure on the $CeIn_3$ crystal,[40] indicating a stronger hybridization strength in $CeMIn_5$ compounds than in $CeIn_3$, which is consistent with our ARPES and STM results.

In summary, we have performed on-resonant ARPES and STM/STS measurements on $CeIn_3$. We find a weakly dispersive quasiparticle band with an energy dispersion of 4 meV near $E_F$, indicating the hybridization between $f$ electrons and conduction electrons. The hybridization is further confirmed by STM/STS results. Moreover, the *c-f* hybridization strength in $CeIn_3$ is weaker than that in the two-dimensional compounds $CeCoIn_5$ and $CeIrIn_5$. These results demonstrate the weak *c-f* hybridization strength at low temperature and shed new light on the transport anomaly in $CeIn_3$.

# ACKNOWLEDGMENTS

This work was supported by the National Natural Science Foundation of China (No. 11504342, 11774320, U1630248), Science Challenge Project (No. TZ2016004), the National Key Research and Development Program of China (No. 2017YFA0303104, No. 2016YFA0300200) and the Dean Foundation of China Academy of Engineering Physics (No. 201501040).